\documentclass{aastex}%
\usepackage{amssymb}
\usepackage{amsmath}
\usepackage{amsfonts}
\usepackage{graphicx}%
\begin{document}
{\LARGE \ \ \ \ \ \ \ \ \ Back-reaction instabilities of relativistic cosmic
rays }

\bigskip

\ \ \ \ \ \ \ \ \ \ \ \ \ \ \ \ \ \ \ \ \ \ \ \ \ \ \ \ \ \ \ \ \ \ \ \ \ \ \ \ \ \ \ \ {\large \ A
K Nekrasov\ }

Institute of Physics of the Earth, Russian Academy of Sciences, 123995 Moscow, Russia;

E-mails: anekrasov@ifz.ru, nekrasov.anatoly@gmail.com

\bigskip

\textbf{ABSTRACT}

We explore streaming instabilities of the electron-ion plasma with
relativistic and ultra-relativistic cosmic rays in the background magnetic
field in the multi-fluid approach. Cosmic rays can be both electrons and ions.
The drift speed of cosmic rays is directed along the magnetic field. In
equilibrium, the return current of the background plasma is taken into
account. One-dimensional perturbations parallel to the magnetic field are
considered. The dispersion relations are derived for transverse and
longitudinal perturbations. It is shown that the back-reaction of magnetized
cosmic rays generates new instabilities one of which has the growth rate that
can approach the growth rate of the Bell instability. These new instabilities
can be stronger than the cyclotron resonance instability. For unmagnetized
cosmic rays, the growth rate is analogous to the Bell one. We compare two
models of the plasma return current in equilibrium with three and four charged
components. Some difference between these models is demonstrated. For
longitudinal perturbations, an instability is found in the case of
ultra-relativistic cosmic rays. The results obtained can be applied to
investigation of astrophysical objects such as the shocks by supernova
remnants, galaxy clusters, intracluster medium and so on, where interaction of
cosmic rays with turbulence of the electron-ion plasma produced by them is of
a great importance for the cosmic-ray evolution.

\bigskip

\section{Introduction}

It is known for a long time that a return current arises in a plasma
penetrated by an external beam current [1]. A theory of this phenomenon for
the laboratory plasma has been developed in a number of early papers [1-5]. It
was shown that the induced plasma current depends on the spatio-temporal shape
of the imposed current and is transferred by plasma species. For external
currents of cylindrical shape, it has been found that the return current is
nearly equal to the imposed beam current and lies almost entirely within the
beam channel [2-4]. In bound magnetized plasma with given nonstationary sheet
current, the return current can change with time and be not equal to the
external current [5]. However, inclusion of the surface current in the
perfectly conducting walls results in the full compensation of both currents [5].

The return currents in astrophysics are considered for media where cosmic rays
are present. It is assumed that in equilibrium the total current of cosmic
rays and plasma is equal to zero. Models are explored, in which the
equilibrium current is directed along [6-8] and across [9, 10] the background
magnetic field. In the case of currents parallel to the magnetic field, one
considers a three-component medium, consisting of the electrons, ions and
cosmic rays, where each species has its own drift velocity [6], as well as a
four-component one. In the last case, one assumes that the background plasma
has no drift velocities while cosmic rays and an additional electron component
(for the proton cosmic rays) drift together [7, 8, 11].

The kinetic consideration of cosmic rays drifting along the magnetic field has
been provided by Achterberg [6], Zweibel [7], Bell [8] and Reville et al. [12]
also for perturbations parallel to the magnetic field. The well-known
non-resonant Bell instability [8] has a large growth rate for perturbation
wavelengths shorter than the mean Larmor radius of cosmic-ray protons defined
by their longitudinal momentum. In this case, the contribution of cosmic rays
to the kinetic dispersion relation is small [7, 8] and the instability is due
to the electron return current. Thus, the back-reaction of cosmic rays is
absent in unstable short-wavelength perturbations mentioned above. In the
opposite case of long-wavelength perturbations, the perturbed currents of
cosmic rays (protons) and of additional electrons compensate each other, if
only the perturbed electric drift of particles is taken into account [7].

However, involving the Doppler-shifted polarizational current (back-reaction)
of cosmic rays is also important for cosmic-ray streaming instabilities. This
effect was not considered in [7, 8]. As we show here, the back-reaction of
magnetized cosmic rays gives rise to new streaming instabilities, one of which
has the growth rate of the order of that of the Bell instability [8] in the
vicinity of the instability threshold and less far from it. However, in the
long-wavelength spectral part, for example, these new instabilities can be
more powerful in comparison with the cyclotron resonance instability.

In the present paper, we investigate streaming instabilities of the
electron-ion plasma in the background magnetic field with cosmic rays up to
ultra-relativistic energies using the multi-fluid approach. We assume that
cosmic rays, which can be both protons and electrons, drift along the magnetic
field. One-dimensional perturbations also parallel to the magnetic field are
treated. In this case, transverse and longitudinal movements are split. For
generality, we take into account the thermal energy exchange between
background electrons and ions and the electron thermal conductivity. We derive
dispersion relations for the transverse and longitudinal perturbations. For
the first case, two models with three and four components described above are
used and corresponding results are compared. (Analogous consideration of these
models for shocks has been provided by Amato and Blasi [13]). New
instabilities due to the back-reaction of relativistic cosmic rays are found.

The paper is organized as follows. Section 2 contains the fundamental
equations for plasma, cosmic rays and electromagnetic fields. Equilibrium
state is discussed in section 3. In section 4, the transverse perturbations
with magnetized and unmagnetized cosmic rays are explored. We investigate
longitudinal perturbations in section 5. In section 6, we discuss results
obtained in the preceding sections. Conclusive remarks are given in section 7.

\bigskip

\section{Basic\ equations for plasma and cosmic rays}

The fundamental equations for the plasma that we consider here are the
following:%
\begin{equation}
\frac{\partial\mathbf{v}_{j}}{\partial t}+\mathbf{v}_{j}\cdot\mathbf{\nabla
v}_{j}=-\frac{\mathbf{\nabla}p_{j}}{m_{j}n_{j}}+\frac{q_{j}}{m_{j}}%
\mathbf{E+}\frac{q_{j}}{m_{j}c}\mathbf{v}_{j}\times\mathbf{B+C}_{j},
\end{equation}
the equation of motion,%
\begin{equation}
\frac{\partial n_{j}}{\partial t}+\mathbf{\nabla}\cdot n_{j}\mathbf{v}_{j}=0,
\end{equation}
the continuity equation,%
\begin{equation}
\frac{\partial T_{i}}{\partial t}+\mathbf{v}_{i}\cdot\mathbf{\nabla}%
T_{i}+\left(  \gamma-1\right)  T_{i}\mathbf{\nabla}\cdot\mathbf{v}_{i}%
=\nu_{ie}^{\varepsilon}\left(  n_{e},T_{e}\right)  \left(  T_{e}-T_{i}\right)
\end{equation}
and%
\begin{equation}
\frac{\partial T_{e}}{\partial t}+\mathbf{v}_{e}\cdot\mathbf{\nabla}%
T_{e}+\left(  \gamma-1\right)  T_{e}\mathbf{\nabla}\cdot\mathbf{v}%
_{e}=-\left(  \gamma-1\right)  \frac{1}{n_{e}}\mathbf{\nabla\cdot q}_{e}%
-\nu_{ei}^{\varepsilon}\left(  n_{i},T_{e}\right)  \left(  T_{e}-T_{i}\right)
\end{equation}
are the temperature equations for ions and electrons. In equations (1) and
(2), the subscript $j=i,e$ denotes the ions and electrons, respectively.
Notations in equations (1)-(4) are the following: $q_{j}$ and $m_{j}$ are the
charge and mass of species $j=i,e$; $\mathbf{v}_{j}$ is the hydrodynamic
velocity; $n_{j}$ is the number density; the terms $\mathbf{C}_{e}=-\nu
_{ei}\left(  \mathbf{v}_{e}-\mathbf{v}_{i}\right)  $ and $\mathbf{C}_{i}%
=-\nu_{ie}\left(  \mathbf{v}_{i}-\mathbf{v}_{e}\right)  $ take into account
the collisional momentum exchange between electrons and ions, where $\nu_{ei}%
$($\nu_{ie}$) is the electron(ion)-ion(electron) collision frequency;
$p_{j}=n_{j}T_{j}$ is the thermal pressure; $T_{j}$ is the temperature;
$\nu_{ie}^{\varepsilon}(n_{e},T_{e})$ ($\nu_{ei}^{\varepsilon}\left(
n_{i},T_{e}\right)  $) is the frequency of the thermal energy exchange between
ions (electrons) and electrons (ions) being $\nu_{ie}^{\varepsilon}%
(n_{e},T_{e})=2\nu_{ie}$ [14]; $n_{i}\nu_{ie}^{\varepsilon}\left(  n_{e}%
,T_{e}\right)  =n_{e}\nu_{ei}^{\varepsilon}\left(  n_{i},T_{e}\right)  $;
$\gamma$ is the ratio of the specific heats; $\mathbf{E}$\textbf{\ }and
$\mathbf{B}$ are the electric and magnetic fields and $c$ is the speed of
light in a vacuum. We include the thermal exchange between electrons and ions
because $\nu_{ie}^{\varepsilon}\left(  n_{e},T_{e}\right)  $($\nu
_{ei}^{\varepsilon}\left(  n_{i},T_{e}\right)  $) must be compared with the
dynamical frequency. The value $\mathbf{q}_{e}$ in equation (4) is the
electron heat flux [14]. In a weakly collisional plasma which is here
considered, the electron Larmor radius is much smaller than the electron
collisional mean free path. In this case, the electron thermal flux is mainly
directed along the magnetic field,
\begin{equation}
\mathbf{q}_{e}=-\chi_{e}\mathbf{b}\left(  \mathbf{b\cdot\nabla}\right)  T_{e},
\end{equation}
where $\chi_{e}$ is the electron thermal conductivity coefficient and
$\mathbf{b=B/}B$ is the unit vector along the magnetic field. We assume that
the thermal flux in equilibrium is absent.

Equations for relativistic cosmic rays we take in the form [15]
\begin{equation}
\frac{\partial\left(  R_{cr}\mathbf{p}_{cr}\right)  }{\partial t}%
+\mathbf{v}_{cr}\cdot\mathbf{\nabla}\left(  R_{cr}\mathbf{p}_{cr}\right)
=-\frac{\mathbf{\nabla}p_{cr}}{n_{cr}}+q_{cr}\left(  \mathbf{E+}\frac{1}%
{c}\mathbf{v}_{cr}\times\mathbf{B}\right)  ,
\end{equation}%
\begin{equation}
\left(  \frac{\partial}{\partial t}+\mathbf{v}_{cr}\cdot\mathbf{\nabla
}\right)  \left(  \frac{p_{cr}\gamma_{cr}^{\Gamma_{cr}}}{n_{cr}^{\Gamma_{cr}}%
}\right)  =0,
\end{equation}
where%
\begin{equation}
R_{cr}=1+\frac{\Gamma_{cr}}{\Gamma_{cr}-1}\frac{T_{cr}}{m_{cr}c^{2}}.
\end{equation}
In these equations, $\mathbf{p}_{cr}=\gamma_{cr}m_{cr}\mathbf{v}_{cr}$ is the
momentum of a cosmic-ray particle having the rest mass $m_{cr}$ and velocity
$\mathbf{v}_{cr}$; $q_{cr}$ is the charge; $p_{cr}=\gamma_{cr}^{-1}%
n_{cr}T_{cr}$ is the kinetic pressure; $n_{cr}$ is the number density in the
laboratory frame; $\Gamma_{cr}$ is the adiabatic index; $\gamma_{cr}=\left(
1-\mathbf{v}_{cr}^{2}/c^{2}\right)  ^{-1/2}$ is the relativistic factor. The
continuity equation is the same as equation (2) for $j=cr$. Equation (8) can
be used for both cold non-relativistic, $T_{cr}\ll$ $m_{cr}c^{2}$, and hot
relativistic, $T_{cr}\gg$ $m_{cr}c^{2}$, cosmic rays. In the first (second)
case, we have $\Gamma_{cr}=5/3$ ($4/3$) [15]. The general form of the value
$R_{cr}$ applying at any relations between $T_{cr}$ and $m_{cr}c^{2}$, can be
found e.g. in [16, 17].

Equations (1)-(4), (6) and (7) are solved together with Maxwell's equations
\begin{equation}
\mathbf{\nabla\times E=-}\frac{1}{c}\frac{\partial\mathbf{B}}{\partial t}%
\end{equation}
and
\begin{equation}
\mathbf{\nabla\times B=}\frac{4\pi}{c}\mathbf{j+}\frac{1}{c}\frac
{\partial\mathbf{E}}{\partial t},
\end{equation}
where $\mathbf{j=j}_{pl}+\mathbf{j}_{cr}=\sum_{j}q_{j}n_{j}\mathbf{v}%
_{j}+\mathbf{j}_{cr}$.

\bigskip

\section{Equilibrium state}

We will consider a uniform plasma embedded in the uniform magnetic field
$\mathbf{B}_{0}$ (subscript $0$ here and below denotes background parameters)
directed along the $z$-axis. We assume that in equilibrium the plasma is
penetrated by a uniform beam of cosmic rays having the uniform streaming
velocity $v_{cr0}$ along the $z$-axis. The return plasma current along this
axis compensating the current of cosmic rays is provided by the streaming
velocities of electrons, $v_{e0}$, and ions, $v_{i0}$. The quasi-neutrality is
satisfied due to cosmic-ray charge neutralizaion from the background
environment [18]. Thus, we have two equations in equilibrium
\begin{equation}
q_{e}n_{e0}v_{e0}+q_{i}n_{i0}v_{i0}+q_{cr}n_{cr0}v_{cr0}=0
\end{equation}
and%
\begin{equation}
q_{e}n_{e0}+q_{i}n_{i0}+q_{cr}n_{cr0}=0.
\end{equation}
Such a three-component model corresponds to the one considered by Achterberg
[6]. In papers by Zweibel [7] and Bell [8], a four-component model has been
explored, in which plasma species are immobile and the additional electrons
(in the case of the proton cosmic rays) have the cosmic-ray number density and
drift with the cosmic-ray drift velocity. We show below that there is some
difference between these two models.

\bigskip

\section{Transverse perturbations}

We will treat one-dimensional perturbations along the background magnetic
field. From equations (9) and (10), it is followed that in this case the
transverse and longitudinal perturbations are split. The transverse wave
equations have the form
\begin{align}
c^{2}\left(  \frac{\partial}{\partial t}\right)  ^{-2}\frac{\partial^{2}%
E_{1x}}{\partial z^{2}}  &  =4\pi\left(  \frac{\partial}{\partial t}\right)
^{-1}j_{1x}+E_{1x},\\
c^{2}\left(  \frac{\partial}{\partial t}\right)  ^{-2}\frac{\partial^{2}%
E_{1y}}{\partial z^{2}}  &  =4\pi\left(  \frac{\partial}{\partial t}\right)
^{-1}j_{1y}+E_{1y}.\nonumber
\end{align}
The perturbed currents $j_{1x,y}$ are given by equations (A15), (A16), (B10)
and (B11). Substituting them into equation (13), we obtain
\begin{equation}
\left[  c^{2}\frac{\partial^{2}}{\partial z^{2}}\left(  \frac{\partial
}{\partial t}\right)  ^{-2}-\varepsilon_{xx}-1\right]  ^{2}E_{1x,y}%
+\varepsilon_{xy}^{2}E_{1x,y}=0,
\end{equation}
where%
\begin{align}
\varepsilon_{xx}  &  =\varepsilon_{plxx}+\varepsilon_{crxx},\\
\varepsilon_{xy}  &  =\varepsilon_{plxy}+\varepsilon_{crxy}.\nonumber
\end{align}
For perturbations of the form $\exp\left(  ik_{z}-i\omega t\right)  $, we find
from equation (14) the dispersion relation%
\begin{equation}
\frac{k_{z}^{2}c^{2}}{\omega^{2}}-\varepsilon_{xx}-1=\pm i\varepsilon_{xy}.
\end{equation}
From equation (13), it is followed that the waves have a circular polarization.

\bigskip

\subsection{\textit{Magnetized species}}

We first consider equation (16) in the case, in which all species are
magnetized i.e.%
\begin{align}
\omega_{cj}^{2}  &  \gg D_{tj}^{2},\\
\omega_{ccr}^{2}  &  \gg D_{cr}^{2},\nonumber
\end{align}
where $\omega_{cj,cr}=q_{j,cr}B_{0}/m_{j,cr}c$ is the cyclotron frequency,
$D_{cr}=\gamma_{cr0}R_{cr0}D_{tcr}$, $D_{tj,cr}=-i\omega+ik_{z}v_{j,cr0}$.
Using conditions (17), we calculate the values $\varepsilon_{plxx,y}$ and
$\varepsilon_{crxx,y}$ given by equations (A16) and (B11), respectively, and
substitute them into equation (15). Then from equation (16), we derive the
following dispersion relation:%
\begin{equation}
k_{z}^{2}c^{2}+\sum_{j}\frac{\omega_{pj}^{2}D_{tj}^{2}}{\omega_{cj}^{2}%
}+\gamma_{cr0}R_{cr0}\frac{\omega_{pcr}^{2}D_{tcr}^{2}}{\omega_{ccr}^{2}}=\mp
i\left(  \sum_{j}\frac{\omega_{pj}^{2}D_{tj}}{\omega_{cj}}+\frac{\omega
_{pcr}^{2}D_{tcr}}{\omega_{ccr}}\right)  .
\end{equation}
In equation (18), we have neglected the contribution of the displacement
current and small terms proportional to $D_{tj,cr}^{3}/\omega_{cj,cr}^{3}$.
According to equations (11) and (12), the right-hand side of equation (18) is
equal to zero. Thus, we obtain \ \
\begin{equation}
\alpha_{e}\left(  \omega-k_{z}v_{e0}\right)  ^{2}+\alpha_{i}\left(
\omega-k_{z}v_{i0}\right)  ^{2}+\alpha_{cr}\left(  \omega-k_{z}v_{cr0}\right)
^{2}=k_{z}^{2}c^{2},
\end{equation}
where $\alpha_{j}=\omega_{pj}^{2}/\omega_{cj}^{2}$ and $\alpha_{cr}%
=\gamma_{cr0}R_{cr0}\omega_{pcr}^{2}/\omega_{ccr}^{2}$. The solution of
equation (19) is given by%
\begin{equation}
\omega=\frac{A_{2}}{A_{1}}k_{z}\pm\frac{1}{A_{1}}k_{z}\left(  A_{2}^{2}%
-A_{1}A_{3}\right)  ^{1/2},
\end{equation}
where%
\begin{align}
A_{1}  &  =\alpha_{e}+\alpha_{i}+\alpha_{cr},\\
A_{2}  &  =\alpha_{e}v_{e0}+\alpha_{i}v_{i0}+\alpha_{cr}v_{cr0},\nonumber\\
A_{3}  &  =\alpha_{e}v_{e0}^{2}+\alpha_{i}v_{i0}^{2}+\alpha_{cr}v_{cr0}%
^{2}-c^{2}.\nonumber
\end{align}
Using equation (21), we find expression $A_{2}^{2}-A_{1}A_{3}$%
\begin{equation}
A_{2}^{2}-A_{1}A_{3}=A_{1}c^{2}-\alpha_{e}\alpha_{i}\left(  v_{e0}%
-v_{i0}\right)  ^{2}-\alpha_{e}\alpha_{cr}\left(  v_{cr0}-v_{e0}\right)
^{2}-\alpha_{i}\alpha_{cr}\left(  v_{cr0}-v_{i0}\right)  ^{2}.
\end{equation}
Equation (20) describes the streaming instability if $\left(  A_{2}^{2}%
-A_{1}A_{3}\right)  <0$.

The number density of cosmic rays is considerably smaller than the number
density of the background plasma. Therefore, we can conclude from equation
(11) that $v_{e,i0}\ll v_{cr0}$. In this case, equation (22) can be written in
the form
\begin{equation}
A_{2}^{2}-A_{1}A_{3}\simeq\left(  \alpha_{i}+\alpha_{cr}\right)  c^{2}%
-\alpha_{i}\alpha_{cr}v_{cr0}^{2}.
\end{equation}
The growth rate of instability $\delta=\operatorname{Im}\omega$ found from
equation (20) in the case $\alpha_{i}\alpha_{cr}v_{cr0}^{2}\gg\left(
\alpha_{i}+\alpha_{cr}\right)  c^{2}$ is equal to%
\begin{equation}
\delta=\frac{\left(  \alpha_{i}\alpha_{cr}\right)  ^{1/2}}{\alpha_{i}%
+\alpha_{cr}}k_{z}v_{cr0}.
\end{equation}
This new instability arises due to the cosmic-ray back-reaction, i.e. due to
the same dynamics of cosmic rays as that of the plasma connected with the
polarizational drift of the species (see equation (18)).

For a four-component model consisting of the background ions and electrons
without drift velocities, proton cosmic rays and additional electrons with the
cosmic-ray number density and drift velocity [7], equation (18) has solution
\[
\omega=\frac{\alpha_{cr}}{\alpha_{i}+\alpha_{cr}}k_{z}v_{cr0}\pm\frac{k_{z}%
}{\alpha_{i}+\alpha_{cr}}\left[  -\alpha_{i}\alpha_{cr}v_{cr0}^{2}+\left(
\alpha_{i}+\alpha_{cr}\right)  c^{2}\right]  ^{1/2}.
\]
We see that this solution gives the same growth rate as that given by
expression (24) (see equation (23)). However, the real frequency (or the phase
velocity) for a four-component model is different from that for a
three-component one.

\bigskip

\subsection{\textit{Unmagnetized cosmic rays}}

In this section, we assume that cosmic rays are unmagnetized%
\begin{equation}
D_{cr}^{2}\gg\omega_{ccr}^{2}.
\end{equation}
If $\omega_{ccr}^{2}\gg\left(  \omega-k_{z}v_{cr0}\right)  ^{2}$, this
condition can be satisfied for relativistic cosmic rays for which
$\gamma_{cr0}R_{cr0}\gg1$. Then, we obtain
\begin{align}
\varepsilon_{crxx}  &  =-\frac{\omega_{pcr}^{2}}{\gamma_{cr0}R_{cr0}\omega
^{2}},\\
\varepsilon_{crxy}  &  =-\frac{\omega_{pcr}^{2}\omega_{ccr}}{\gamma
_{cr0}R_{cr0}D_{cr}\omega^{2}}.\nonumber
\end{align}
The plasma ions and electrons stay magnetized. Substituting equation (26) and
$\varepsilon_{plxx,y}$ into equation (16), we will have
\begin{equation}
k_{z}^{2}c^{2}-\alpha_{e}\left(  \omega-k_{z}v_{e0}\right)  ^{2}-\alpha
_{i}\left(  \omega-k_{z}v_{i0}\right)  ^{2}=\pm\beta_{cr}\left(  \omega
-k_{z}v_{cr0}\right)  ,
\end{equation}
where $\beta_{cr}=\omega_{pcr}^{2}/\omega_{ccr}$. When obtaining the
right-hand side of this equation, we have used equations (11) and (12). We
note that equation (27) does not contain the contribution of the cosmic-ray
perturbed dynamics, which is small in a comparison with the plasma current
produced by the electric drift velocities of ions and electrons. Solution of
equation (27) is given by%
\begin{align}
\omega &  =\frac{1}{\alpha_{i}}\left(  \alpha_{e}k_{z}v_{e0}+\alpha_{i}%
k_{z}v_{i0}\mp\frac{\beta_{cr}}{2}\right) \\
&  \pm\frac{1}{\alpha_{i}}\left[  \pm\alpha_{i}\beta_{cr}k_{z}v_{cr0}%
-\alpha_{e}\alpha_{i}k_{z}^{2}\left(  v_{e0}-v_{i0}\right)  ^{2}+\frac{1}%
{4}\beta_{cr}^{2}+\alpha_{i}k_{z}^{2}c^{2}\right]  ^{1/2}.\nonumber
\end{align}
From equations (11) and (12), it is followed that $v_{e0}-v_{i0}\simeq\left(
q_{cr}n_{cr0}/q_{i}n_{i0}\right)  v_{cr0}$ ($q_{i}=-q_{e}$). An estimation of
the ratio of the second term in the squared brackets in equation (28) to the
first one gives the value $\left(  n_{cr0}/n_{i0}\right)  \left(  k_{z}%
v_{cr0}/\omega_{ce}\right)  $, which is generally speaking much smaller than unity.

Solution of the dispersion relation for the four-component medium considered
above is the following:%
\begin{equation}
\omega=\mp\frac{1}{2}\frac{\beta_{cr}}{\alpha_{i}}\pm\frac{1}{\alpha_{i}%
}\left(  \pm\alpha_{i}\beta_{cr}k_{z}v_{cr0}-\alpha_{i}\frac{\beta_{cr}%
}{\left\vert \omega_{ce}\right\vert }k_{z}^{2}v_{cr0}^{2}+\frac{1}{4}%
\beta_{cr}^{2}+\alpha_{i}k_{z}^{2}c^{2}\right)  ^{1/2},
\end{equation}
where the sign $\left\vert {}\right\vert $ denotes an absolute value. We see
some differences between equations (28) and (29). The growth rates are the
same (neglecting the small terms), while the phase velocities are different
for two models.

Equation (29) applied to the proton cosmic rays coincides with equation (8)
given in the paper by Zweibel and Everett [11], if we neglect the term
proportional to $v_{cr0}^{2}$ (assuming that $k_{z}v_{cr0}\ll\left\vert
\omega_{ce}\right\vert $) and take the lower sign (see also [7, 8]). This
coincidence is due to the absence of the dynamical contribution of
unmagnetized cosmic rays to the dispersion relation (27) as it is also in the
case considered in [7, 8, 11]. However, conditions of unmagnetization are
different in both cases. In our one-dimensional magnetohydrodynamic case, the
transverse perturbations of cosmic rays do not contain the thermal pressure,
and the condition of unmagnetization has the form (25). At the same time, the
kinetic consideration shows that cosmic rays are also unmagnetized in
perturbations with wavelengths much smaller than their Larmor radius defined
by the thermal velocity along the magnetic field [7, 8]. Thus, in both
limiting cases, the back-reaction of cosmic rays is negligible that results in
the same growth rates of instability due to the return plasma current.

We note that if we set $v_{e,i0}=0$ in equation (28) (or on the left-hand side
of equation (27)), we return to equation (29) without the term $\sim$
$v_{cr0}^{2}$.

\bigskip

\section{Longitudinal perturbations}

\subsection{\textit{Dispersion relation}}

We now consider potential perturbations along the background magnetic field.
The wave equation is the following (see equation (10)):%

\begin{equation}
4\pi j_{1z}\mathbf{+}\frac{\partial E_{1z}}{\partial t}=0.
\end{equation}
In Appendices A and B, there are obtained the plasma, $j_{pl1z}$, and cosmic
ray, $j_{cr1z}$, perturbed currents (equations (A17) and (B12), respectively).
Substitution them into equation (30) and the Fourier transformation lead to
the dispersion relation%
\begin{equation}
0=\frac{D}{L}\left[  \frac{\omega_{pe}^{2}}{D_{te}}\left(  L_{1i}D_{te}%
-L_{2e}D_{ti}\frac{q_{i}m_{e}}{q_{e}m_{i}}\right)  +\frac{\omega_{pi}^{2}%
}{D_{ti}}\left(  L_{1e}D_{ti}-L_{2i}D_{te}\frac{q_{e}m_{i}}{q_{i}m_{e}%
}\right)  \right]  +\frac{\omega_{pcr}^{2}}{L_{cr}}+1,
\end{equation}
where $\partial/\partial t=-i\omega$ and $\partial/\partial z=ik_{z}$. This
equation will be treated in the limiting cases.

\bigskip

\subsection{\textit{Cold electrons and ions}}

We first consider the cold plasma species for which%
\begin{align}
D_{te}\left(  D_{te}+\nu_{ei0}\right)   &  \gg k_{z}^{2}\frac{T_{0}}{m_{e}},\\
D_{ti}\left(  D_{ti}+\nu_{ie0}\right)   &  \gg k_{z}^{2}\frac{T_{0}}{m_{i}%
}.\nonumber
\end{align}
For cosmic rays, we here and below assume that the following condition is
satisfied:%
\begin{equation}
D_{tcr}^{2}\gg\frac{\Gamma_{cr}T_{cr0}}{\gamma_{cr0}^{4}E_{cr0}m_{cr}}%
k_{z}^{2},
\end{equation}
where
\[
E_{cr0}=R_{cr0}-\frac{\Gamma_{cr}T_{cr0}}{m_{cr}c^{2}}\frac{v_{cr0}^{2}}%
{c^{2}}.
\]
In this case, the first term on the right-hand side of equation (B13) is
dominant. We note that the temperature of cosmic rays can be relativistic,
i.e. $T_{cr0}/m_{cr}c^{2}\gg1$. Then, using equations (A6), (A8), (A10), (A12)
and (B13) under conditions defined by equations (32) and (33), we obtain
equation (31) in the form%
\begin{equation}
0=\frac{1}{\left(  D_{te}D_{ti}+D_{te}\nu_{ie0}+D_{ti}\nu_{ei0}\right)
}\left(  \frac{\omega_{pe}^{2}}{D_{te}}D_{ti}+\frac{\omega_{pi}^{2}}{D_{ti}%
}D_{te}\right)  +\frac{\omega_{pcr}^{2}}{\gamma_{cr0}^{3}E_{cr0}D_{tcr}^{2}},
\end{equation}
where for simplicity we have neglected unity.

\subsubsection{Collisionless case}

We now assume that%
\[
D_{te,i}\gg\nu_{ei,ie0}.
\]
Then equation (34) takes the form%
\begin{equation}
\frac{\omega_{pe}^{2}}{\left(  \omega-k_{z}v_{e0}\right)  ^{2}}+\frac
{\omega_{pi}^{2}}{\left(  \omega-k_{z}v_{i0}\right)  ^{2}}+\frac{\omega
_{pcr}^{2}}{\gamma_{cr0}^{3}E_{cr0}\left(  \omega-k_{z}v_{cr0}\right)  ^{2}%
}=0.
\end{equation}
In the vicinity of $\omega\approx k_{z}v_{cr0}$, when the back-reaction of
cosmic rays plays a role, solution of equation (35) is the following:%
\begin{equation}
\omega=k_{z}v_{cr0}\left(  1+i\gamma_{cr0}^{-3/2}E_{cr0}^{-1/2}\frac
{\omega_{pcr}}{\omega_{pe}}\right)  .
\end{equation}
In the region $\omega\approx k_{z}v_{i0}$, equation (35) gives
\begin{equation}
\omega=k_{z}v_{i0}+i\left(  \frac{m_{e}}{m_{i}}\right)  ^{1/2}k_{z}\left\vert
v_{i0}-v_{e0}\right\vert ,
\end{equation}
where $v_{i0}-v_{e0}\simeq-\left(  n_{cr}/n_{i0}\right)  v_{cr0}$ (see (11)).
The ratio of the growth rate defined by equation (37) to that of equation (36)
is equal to $\gamma_{cr0}^{3}E_{cr0}\left(  m_{cr}/m_{i}\right)  \left(
n_{cr0}/n_{i0}\right)  $. This value can be much less then unity even at
$\gamma_{cr0}\gg1$ and $T_{cr0}\gg$ $m_{cr}c^{2}$.

\subsubsection{Collisional case}

In the collisional case%
\[
\nu_{ei0}\gg D_{te},
\]
we find from equation (34) solution in the region $\omega\approx k_{z}v_{cr0}$%
\begin{equation}
\omega=k_{z}v_{cr0}+\frac{\left(  -1+i\right)  }{\sqrt{2}}\left(  \nu
_{ei0}k_{z}v_{cr0}\right)  ^{1/2}\gamma_{cr0}^{-3/2}E_{cr0}^{-1/2}\frac
{\omega_{pcr}}{\omega_{pe}}.
\end{equation}
Thus, the back-reaction of relativistic cosmic rays can result in an
instability of potential perturbations.

\bigskip

\subsection{\textit{Hot electrons and cold/ hot ions}}

Consideration shows that in the cases $D_{te}\left(  D_{te}+\nu_{ei0}\right)
\ll\left(  T_{0}/m_{e}\right)  k_{z}^{2},D_{ti}\left(  D_{ti}+\nu
_{ie0}\right)  \gg\left(  T_{0}/m_{i}\right)  k_{z}^{2}$ and $D_{ti}\left(
D_{ti}+\nu_{ie0}\right)  \ll\left(  T_{0}/m_{i}\right)  k_{z}^{2}$ the
frequency $\omega$ is of the order of $k_{z}v_{cr0}$ as that in equations (36)
and (40). Equation (33) results in condition $\gamma_{cr0}\gg1$ when
$v_{cr0}\simeq c$. Thus, the temperature of the background plasma should be
relativistic. However, this contradicts the basic equations, where a plasma is
a non-relativistic one. Therefore, conditions for hot plasma are invalid.
Taking into account other terms in equation (B13) does not give an instability.

\bigskip

\section{Discussion}

We now discuss the growth rates and conditions of their derivation for
transverse perturbations considered in section 4. For magnetized cosmic rays
obeying condition (17), the growth rate is given by equation (24). Below, we
assume that ions and cosmic rays are the protons. Let us first consider the
case in which $\alpha_{i}\gg\alpha_{cr}$ or%
\begin{equation}
1\gg\gamma_{cr0}R_{cr0}\frac{n_{cr0}}{n_{i0}}.
\end{equation}
Then, the condition of instability can be written in the "soft" form%
\begin{equation}
\gamma_{cr0}R_{cr0}\frac{n_{cr0}}{n_{i0}}\gtrsim\frac{c_{Ai}^{2}}{v_{cr0}^{2}%
},
\end{equation}
where $c_{Ai}=\left(  B_{0}^{2}/4\pi n_{i0}m_{i}\right)  ^{1/2}$ is the ion
Alfv\'{e}n velocity (see equation (23)). The growth rate is equal to%
\begin{equation}
\delta=\left(  \gamma_{cr0}R_{cr0}\frac{n_{cr0}}{n_{i0}}\right)  ^{1/2}%
k_{z}v_{cr0}.
\end{equation}
This growth rate increases with the wave number $k_{z}$. However, the value
$k_{z}$ is limited from above by condition of magnetization (17). For cosmic
rays, this condition has the form%
\[
\frac{\omega_{ci}^{2}}{\gamma_{cr0}^{2}R_{cr0}^{2}v_{cr0}^{2}}\gg k_{z}^{2}.
\]
If we set, for estimation,
\[
k_{z\max}\sim\frac{\omega_{ci}}{\gamma_{cr0}R_{cr0}v_{cr0}}%
\]
and substitute this value to expression (41), we obtain the maximal growth
rate $\delta_{\max}$%
\[
\delta_{\max}\sim\omega_{ci}\left(  \frac{1}{\gamma_{cr0}R_{cr0}}\frac
{n_{cr0}}{n_{i0}}\right)  ^{1/2}.
\]
We note that according to condition (40), $\delta_{\max}\lesssim
\delta_{\text{Bell}}$ and $k_{z\max}\lesssim k_{\text{Bell}}$, where
\[
\delta_{\text{Bell}}=\frac{1}{2}\omega_{ci}\frac{n_{cr0}}{n_{i0}}\frac
{v_{cr0}}{c_{Ai}}%
\]
and
\[
k_{\text{Bell}}=\frac{1}{2}\omega_{ci}\frac{n_{cr0}}{n_{i0}}\frac{v_{cr0}%
}{c_{Ai}^{2}}%
\]
are the growth rate and the wave number of the fastest growing mode for the
Bell instability [8, 11]. From equations (20) and (21), we see that
$\operatorname{Re}\omega\lesssim\delta$.

The case $\alpha_{cr}\gg\alpha_{i}$ or
\begin{equation}
\gamma_{cr0}R_{cr0}\frac{n_{cr0}}{n_{i0}}\gg1
\end{equation}
can be satisfied for ultra-relativistic cosmic rays for which $\gamma_{cr0}%
\gg1$ and/or $R_{cr0}\gg1$. In the last case, the temperature of cosmic rays
is relativistic one, $T_{cr0}\gg m_{cr}c^{2}$. The "soft" condition of
instability has the form%
\begin{equation}
v_{cr0}^{2}\gtrsim c_{Ai}^{2}.
\end{equation}
The growth rate is equal to
\begin{equation}
\delta=\left(  \frac{1}{\gamma_{cr0}R_{cr0}}\frac{n_{i0}}{n_{cr0}}\right)
^{1/2}k_{z}v_{cr0}.
\end{equation}
In the case under consideration, we have $\operatorname{Re}\omega=k_{z}%
v_{cr0}\gg\delta$ (see equations (20) and (21)). Thus, we find from (17) the
upper limit for $k_{z}^{2}$
\[
\frac{\omega_{ci}^{2}}{v_{cr0}^{2}}\frac{1}{\gamma_{cr0}R_{cr0}}\frac{n_{cr0}%
}{n_{i0}}\gg k_{z}^{2}.
\]
Taking, for estimation,
\[
k_{z\max}=\frac{\omega_{ci}}{v_{cr0}}\left(  \frac{1}{\gamma_{cr0}R_{cr0}%
}\frac{n_{cr0}}{n_{i0}}\right)  ^{1/2}%
\]
and substituting this expression to (44), we find%
\[
\delta_{\max}=\frac{\omega_{ci}}{\gamma_{cr0}R_{cr0}}.
\]
From conditions (42) and (43), it is followed that $\delta_{\max}\ll
\delta_{\text{Bell}}$ and $k_{z\max}\ll k_{\text{Bell}}$.

In the case $\alpha_{i}\sim\alpha_{cr}$, solution (20) takes the form%
\begin{equation}
\omega=\frac{1}{2}\left(  1+i\right)  k_{z}v_{cr0},
\end{equation}
when $v_{cr0}^{2}\gtrsim2c_{Ai}^{2}$. The upper limit for $k_{z}$ for solution
(45) is the same as for the case (39). Thus,%
\[
\delta_{\max}=\frac{1}{2}\omega_{ci}\frac{n_{cr0}}{n_{i0}}.
\]

Let us compare, for example, solution (41) with the growth rate $\delta_{res}$
of the cyclotron resonance instability of cosmic rays [29], which is thought
to play a crucial role in the early stages of cosmic-ray acceleration in
shocks (e.g., [30]). For the real frequency $\omega=k_{z}c_{Ai}$, the growth
rate $\delta_{res}\ll k_{z}c_{Ai}$ for a particular distribution function [7]
can be written in the form
\begin{equation}
\delta_{res}=\frac{1}{2}\omega_{ci}\frac{n_{cr0}}{n_{i0}}\left(  \frac
{v_{cr0}}{c_{Ai}}-1\right)  \frac{p_{1}/p_{0}}{1+\left(  p_{1}/p_{0}\right)
^{2}},
\end{equation}
where $p_{1}=m_{cr}\omega_{ccr}/k_{z}$ and $p_{0}$ is a typical momentum of
cosmic rays. This growth rate has a maximum of the order of the Bell growth
rate at $p_{1}=p_{0}$, when a wavelength of perturbation is equal to a typical
Larmor radius $\rho_{cr}=p_{0}/m_{cr}\omega_{ccr}$ multiplied by $2\pi$. In
the long-wavelength part of spectrum, $p_{1}\gg p_{0}$ or $1\gg k_{z}\rho
_{cr}$, expression (46) becomes
\begin{equation}
\delta_{res}=\frac{1}{2}\omega_{ci}\frac{n_{cr0}}{n_{i0}}\left(  \frac
{v_{cr0}}{c_{Ai}}-1\right)  k_{z}\rho_{cr}.
\end{equation}
The ratio of the growth rate (41) to that of (47) is equal to%
\[
\frac{\delta}{\delta_{res}}\approx2\left(  \gamma_{cr0}R_{cr0}\frac{n_{i0}%
}{n_{cr0}}\right)  ^{1/2}\frac{m_{cr}c_{Ai}}{p_{0}}.
\]
The case $\delta\gg\delta_{res}$ results in
\begin{equation}
4\gamma_{cr0}R_{cr0}\frac{n_{i0}}{n_{cr0}}\gg\frac{p_{0}^{2}}{m_{cr}^{2}%
c_{Ai}^{2}}.
\end{equation}
Condition (48) can be satisfied. An analogous consideration for solution (44)
gives%
\[
\frac{\delta}{\delta_{res}}=2\left(  \frac{1}{\gamma_{cr0}R_{cr0}}\frac
{n_{i0}^{3}}{n_{cr0}^{3}}\right)  ^{1/2}\frac{m_{cr}c_{Ai}}{p_{0}}.
\]
In this case, it is also possible to be\textbf{ }$\delta\gg\delta_{res}%
$.\textbf{ }

For unmagnetized cosmic rays satisfying condition (25) and magnetized
background electrons and ions, solution of the equation (27) is given by
equation (28). In the case $v_{cr0}^{2}>c_{Ai}^{2}$ to neglect the term
$\beta_{cr}^{2}$, the growth rate has the form $\delta_{\text{Bell}}$ (see
above). The frequency $\omega$ is smaller than $k_{\text{Bell}}v_{cr0}$. Thus,
condition (25) takes the form%
\begin{equation}
\gamma_{cr0}R_{cr0}\frac{n_{cr0}}{n_{i0}}\frac{v_{cr0}^{2}}{c_{Ai}^{2}}\gg1,
\end{equation}
where we have inserted $k_{\text{Bell}}$. We note that under condition (49)
cosmic rays do not contribute to the dispersion relation, i.e. the cosmic-ray
back-reaction is absent. In the kinetic consideration, we obtain an analogous
result for hot cosmic rays $p_{\Vert cr}\gg m_{cr}\omega_{ccr}/k_{z}$, where
$p_{\Vert cr}$ is the average momentum along the magnetic field [7, 8].
Substitution to the last condition of the value $k_{\text{Bell}}$ gives
\[
\frac{n_{cr0}}{n_{i0}}\frac{v_{cr0}}{c_{Ai}^{2}}\frac{p_{\Vert cr}}{m_{cr}}%
\gg1.
\]

Let us discuss longitudinal perturbations. In the case of the cold background
plasma expressed by condition (32) and at condition (33) for cosmic rays,
solution of equation (34) in the collisionless case is given by equation (36).
Condition (32) can be written as $v_{cr0}^{2}\gg T_{0}/m_{e}$. For cosmic
rays, condition (33) takes the form\ \
\begin{equation}
\frac{n_{cr0}}{n_{e0}}\gg\frac{\Gamma_{cr}T_{cr0}}{\gamma_{cr0}m_{e}%
v_{cr0}^{2}}.
\end{equation}
Condition (50) can be satisfied for ultra-relativistic cosmic rays with
$\gamma_{cr0}\gg1$ when $v_{cr0}\simeq c$. The growth rate $\delta$ is the
following:%
\begin{equation}
\delta=\gamma_{cr0}^{-3/2}E_{cr0}^{-1/2}\left(  \frac{m_{e}}{m_{cr}}%
\frac{n_{cr0}}{n_{e0}}\right)  ^{1/2}k_{z}v_{cr0}.
\end{equation}
This growth rate is considerably smaller than that given, for example, by
equation (41). The collisional growth rate (38) is larger than the
collisionless one (51) by a factor of $\left(  \nu_{ei0}/k_{z}v_{cr0}\right)
^{1/2}\gg1$\textbf{.}

In our investigation, we have not included collisions in the momentum equation
for the transverse perturbations. It can be shown that in the present case it
is possible under condition $\omega_{ci}^{2}\gg\nu_{ie}D_{ti}^{2}/\omega$
(e.g., [31, 32]) (see also (A15) and (A16)). In the temperature equations, we
did not take into account the heating due to viscosity and the Joule heating.
These effects can result, in particular, in damping of perturbations [14, 29].
For our model, the resistive damping $\delta_{\text{Joule}}$ is equal to
$\delta_{\text{Joule}}=k_{z}^{2}c^{2}/8\pi\sigma$, where $\sigma=q_{e}%
^{2}n_{e0}/\nu_{ei}m_{e}$ is the electric conductivity, and the viscous
damping is $\delta_{\text{visc}}=0.6k_{z}^{2}T_{i0}\nu_{ii}/\omega_{ci}%
^{2}m_{i}$, where $\nu_{ii}$ is the ion-ion collision frequency [14, 29]. In
the paper [29], it has been shown that these dampings are negligible in
comparison, for example, with the ion-neutral collision damping. We here also
assume that the growth rates can exceed the dissipative effects. The presence
of the background plasma current in equilibrium can also give rise to other
specific instabilities (e.g., [33]). However, all these additional questions
are out of the scope of the present paper, which is devoted to effect of
back-reaction of streaming cosmic rays.

The streaming instabilities driven by cosmic rays may play a significant role
in such environments as the shocks caused by supernova remnants [8, 19-22],
galaxy clusters [23, 24], intracluster medium [25-28] and so on, where weakly
collisional plasma consists mainly of electrons and ions (protons) and where
high-energy cosmic rays are present. Therefore, our model and results are
applicable to these astrophysical objects. The main point of this
investigation is finding that the back-reaction of magnetized cosmic rays can
give rise to instabilities, the growth rate of which can approach to that
obtained earlier [8]. Although, the kinetic derivation of the dispersion
relation in [7, 8] contains the dynamics of cosmic rays, the contribution of
the latter to the dispersion relation is negligible in the hot regime. The
same result is obtained for unmagnetized cosmic rays in the fluid
approximation. In these cases, instabilities arise due to the return plasma current.

The exploration carried out in this paper is relevant to the problem of
generation of magnetic fields and acceleration of high-energy cosmic rays. At
present, it is assumed that acceleration of cosmic rays occurs in supernova
remnant shocks due to their multiple crossing of the shock front. This
mechanism is known as the first order Fermi acceleration. The multiple
crossing happens as a result of cosmic-ray diffusion on magnetic
inhomogeneities in the upstream and downstream regions of the shock being
generated by possible instabilities. Such a process in a whole is called the
diffusive shock acceleration [34-38]. One powerful streaming instability has
been found by Bell [8], where the unperturbed cosmic-ray current was directed
along the magnetic field. In the perturbed state, cosmic rays have been
considered as unmagnetized with the Larmor radius defined by the longitudinal
velocity much larger than wavelengths of perturbations. In the nonlinear
regime, this instability amplifies magnetic fields in the upstream medium of
shocks by a factor up to $\sim10$ larger than typically expected in the
interstellar medium [39]. However, X-ray observations [40, 41] show that
magnetic fields in the downstream medium are $\sim100$ times larger than in
the interstellar medium. Therefore, the search for new instabilities has been
continued. One possibility using the pre-amplified magnetic fields by the Bell
instability has been discussed by Riquelme and Spitkovsky [9]. In this paper,
it has been shown that a new instability can arise due to the background
cosmic-ray current streaming across the background magnetic field. The growth
rate of the same order of magnitude as for the Bell instability has been
found. However, Riquelme and Spitkovsky [9] have not considered the
back-reaction of cosmic rays in their analytical treatment. In the paper by
Nekrasov and Shadmehri [10], we have included the back-reaction of cosmic rays
in the multi-fluid approach for the model considered by Riquelme and
Spitkovsky [9] and found a growth rate for the streaming instability
considerably larger than that of Bell [8] and of Riquelmi and Spitkovsky [9]
amounting to the square root of the ratio of plasma to cosmic-ray number
densities. Therefore, it was of interest to take into account this effect also
for the model considered by Bell [8]. For magnetized cosmic rays, we have
found new instabilities, one of which has the growth rate comparable to that
of Bell in the vicinity of the threshold of instability and smaller far from
it in the wavelength region $k_{z}\lesssim k_{\text{Bell}}$. Another
instability for ultra-relativistic cosmic rays is weaker than the Bell one and
excites at $k_{z}\ll k_{\text{Bell}}$. Thus, magnetized cosmic rays can also
amplify magnetic fields, which results in their diffusion in astrophysical
settings. In shock wave fronts, these additional magnetic perturbations will
increase the diffusion of cosmic rays and accordingly the efficiency of their
acceleration. We have shown in our model that electrostatic perturbations can
also be excited by streaming cosmic rays. 

The results obtained represent a contribution to the picture of cosmic-ray
acceleration studied in previous investigations and of generation of magnetic
fields in other astrophysical objects. Taking into account the cosmic-ray
back-reaction can be done by making use of the multi-fluid approach, in which
all species have their own velocities. 

\bigskip

\section{Conclusion}

Using the multi-fluid approach, we have investigated streaming instabilities
of the magnetized electron-ion plasma with relativistic and ultra-relativistic
cosmic rays. Cosmic rays have been assumed to drift along the background
magnetic field. The return current of the background plasma in equilibrium has
been taken into account. One-dimensional perturbations parallel to the
magnetic field have been considered. We have derived dispersion relations for
the transverse and longitudinal perturbations, whose electric field is
polarized across and along the magnetic field, respectively. We have shown
that the back-reaction of magnetized cosmic rays in transverse perturbations
can result in new instabilities, one of which has the growth rate of the order
of that of the Bell instability [8] in the vicinity of the instability
threshold and less far from it. However, in the long-wavelength spectral part,
for example, these new instabilities can be more powerful in comparison with
the cyclotron resonance instability. For unmagnetized cosmic rays, we have
obtained the growth rate, which is the same as the Bell one. For longitudinal
perturbations, we have found an instability in the case of ultra-relativistic
cosmic rays. The corresponding growth rate is less than that for transverse perturbations.

The results obtained can be applied to investigation of astrophysical objects
such as supernova remnant shocks, galaxy clusters, intracluster medium, and so
on, where interaction of cosmic rays with turbulence of the electron-ion
plasma produced by them is of a great importance for the cosmic-ray scattering
and acceleration.

\bigskip

\textbf{Acknowledgements} I gratefully thank both anonymous referees for their
very constructive and useful comments which have helped considerably to
improve this paper.

\bigskip

\bigskip

\section*{{\textbf{References}}}

\bigskip

[1] Roberts T G and Bennett W H 1968 \textit{Plasma Phys.} \textbf{10} 381

[2] Cox J L, Jr. and Bennett W H 1970 \textit{Phys. Fluids} \textbf{13} 182

[3] Hammer D A and Rostoker N 1970 \textit{Phys. Fluids} \textbf{13} 1831

[4] Lee R and Sudan R N 1971 \textit{Phys. Fluids} \textbf{14} 1213

[5] Berk H L and Pearlstein L D 1976 \textit{Phys. Fluids} \textbf{19} 1831

[6] Achterberg A 1983 \textit{Astron. Astrophys. }\textbf{119} 274

[7] Zweibel E G 2003 \textit{Astrophys. J. }\textbf{587 }625

[8] Bell A R 2004 \textit{Mon. Not. R. Astron. Soc.} \textbf{353} 550

[9] Riquelme M A and Spitkovsky A 2010 \textit{Astrophys. J. }\textbf{717} 1054

[10] Nekrasov A K and Shadmehri M 2012 \textit{Astrophys. J.} \textbf{756}
77\textit{ \ }

[11] Zweibel E G and Everett J E 2010 \textit{Astrophys. J. }\textbf{709 }1412

[12] Reville B, Kirk J G, Duffy P and O'Sullivan S 2007 \textit{Astron.
Astrophys. }\textbf{475} 435

[13] Amato E and Blasi P 2009 \textit{Mon. Not. R. Astron. Soc.} \textbf{392} 1591

[14] Braginskii S I 1965 \textit{Rev. Plasma Phys. }\textbf{1} 205

[15] Lontano M Bulanov S and Koga J 2001 \textit{Phys. Plasmas} \textbf{8 }5113

[16] Toepfer A J 1971 \textit{Phys. Rev. A} \textbf{3 }1444

[17] Dzhavakhishvili D I and Tsintsadze N L 1973 \textit{Sov. Phys. JETP
}\textbf{37 }666; 1973 \textit{Zh. Eksp. Teor. Fiz. }\textbf{64 }1314

[18] Alfv\'{e}n H 1939 \textit{Phys. Rev.} \textbf{55} 425

[19] Koyama K, Petre R, Gotthelf E V, Hwang U, Matsuura M, Ozaki M and Holt S
S 1995 \textit{Nature} \textbf{378} 255

[20] Allen G E et al. 1997 \textit{Astrophys. J. }\textbf{487 }L97

[21] Tanimori T et al. 1998 \textit{Astrophys. J. }\textbf{497 }L25

[22] Vink J and Laming J M 2003 \textit{Astrophys. J. }\textbf{584 }758

[23] Brunetti G, Setti G, Feretti L and Giovannini G 2001 \textit{Mon. Not. R.
Astron. Soc. }\textbf{320 }365

[24] Pfrommer C and En\ss lin T A 2004 \textit{Astron. Astrophys.
}\textbf{413} 17

[25] En\ss lin T A 2003 \textit{Astron. Astrophys. }\textbf{399} 409

[26] Guo F and Oh S P 2008 \textit{Mon. Not. R. Astron. Soc. }\textbf{384} 251

[27] Sharma P, Chandran B D G, Quataert E and Parrish I J 2009
\textit{Astrophys. J. }\textbf{699 }348

[28] Sharma P, Parrish I J and Quataert E 2010 \textit{Astrophys. J.
}\textbf{720 }652

[29] Kulsrud R and Pearce W P 1969 \textit{Astrophys. J. }\textbf{156} 445

[30] Gargat\'{e} L and Spitkovsky A 2012 \textit{Astrophys. J.} \textbf{744} 67

[31] Nekrasov A K and Shadmehri M 2010 \textit{Astrophys . J.} \textbf{724 }1165

[32] Nekrasov A K 2011 \textit{Astrophys. J.} \textbf{739 }88\textit{
\textbf{\ }}

[33] Heyvaerts J 1974 \textit{Astron. Astrophys.} \textbf{37} 65\ \textit{ }

[34] Krymskii G F 1977 \textit{Sov. Phys. Dokl}. \textbf{22} 327

[35] Axford W I, Leer E and Skadron G 1977 in \textit{Int. Cosmic Ray Conf.}
\textbf{11 }132

[36] Bell A R 1978 \textit{Mon. Not. R. Astron. Soc. }\textbf{182 }147

[37] Blandford R D and Ostriker J P 1978 \textit{Astrophys. J.} \textbf{221} L29

[38] Drury L O'c 1983 \textit{Rep. Prog. Phys}. \textbf{46 }973

[39] Riquelme M A and Spitkovsky A 2009 \textit{Astrophys. J. }\textbf{694 }626

[40] Ballet J 2006 \textit{Adv. Space Res. }\textbf{37} 1902

[41] Uchiyama Y, Aharonian F A, Tanaka T, Takahashi T and Maeda Y 2007
\textit{Nature} \textbf{449} 576

\bigskip

\begin{appendix}
\section{Appendix}
\subsection{Perturbed velocities of ions and electrons}
Let us put in equation (1) $\mathbf{v}_{j}=\mathbf{v}_{j0}+\mathbf{v}_{j1}$,
$p_{j}=p_{j0}+p_{j1}$, $\mathbf{E=E}_{0}+\mathbf{E}_{1}$, $\mathbf{B=B}%
_{0}+\mathbf{B}_{1}$. We assume that the medium and background velocities of
species are uniform. Then for perturbations depending only on the
$z$-coordinate and $\mathbf{v}_{j0}\parallel\mathbf{z}$, where $\mathbf{z}$ is
the unit vector along the $z$-axis, the linearized equation (1) takes the form%
\begin{equation}
D_{tj}\mathbf{v}_{j1}=-\frac{\mathbf{\nabla}T_{j1}}{m_{j}}-\frac
{T_{j0}\mathbf{\nabla}n_{j1}}{m_{j}n_{j0}}+\mathbf{F}_{j1}\mathbf{+}%
\frac{q_{j}}{m_{j}c}\mathbf{v}_{j1}\times\mathbf{B}_{0}, \tag{A1}%
\end{equation}
where we have used that $p_{j1}=n_{j0}T_{j1}+n_{j1}T_{j0}$ ($n_{j}%
=n_{j0}+n_{j1}$, $T_{j}=T_{j0}+T_{j1}$) and introduced the notations
$D_{tj}=\partial/\partial t+v_{j0}\partial/\partial z$ and%
\begin{align}
\mathbf{F}_{i1}  &  =\frac{q_{i}}{m_{i}}\mathbf{E}_{1}\mathbf{+}\frac{q_{i}%
}{m_{i}c}\mathbf{v}_{i0}\times\mathbf{B}_{1}+\nu_{ie0}\frac{\partial v_{e1z}%
}{D_{te}\partial z}\left(  \mathbf{v}_{i0}-\mathbf{v}_{e0}\right)  -\nu
_{ie0}\left(  \mathbf{v}_{i1z}-\mathbf{v}_{e1z}\right)  ,\tag{A2}\\
\mathbf{F}_{e1}  &  =\frac{q_{e}}{m_{e}}\mathbf{E}_{1}\mathbf{+}\frac{q_{e}%
}{m_{e}c}\mathbf{v}_{e0}\times\mathbf{B}_{1}+\nu_{ei0}\frac{\partial v_{i1z}%
}{D_{ti}\partial z}\left(  \mathbf{v}_{e0}-\mathbf{v}_{i0}\right)  -\nu
_{ei0}\left(  \mathbf{v}_{e1z}-\mathbf{v}_{i1z}\right)  .\nonumber
\end{align}
We do not include collisions for transverse perturbations. The corresponding
condition is given in Discussion. When obtaining (A2), we have taken into
account a perturbation of collision frequency and used equation (2).
From equation (A1), we find equations for the perturbed transverse velocities
$v_{j1x,y}$%
\begin{align}
\left(  D_{tj}^{2}+\omega_{cj}^{2}\right)  v_{j1x}  &  =\omega_{cj}%
F_{j1y}+D_{tj}F_{j1x},\tag{A3}\\
\left(  D_{tj}^{2}+\omega_{cj}^{2}\right)  v_{j1y}  &  =-\omega_{cj}%
F_{j1x}+D_{tj}F_{j1y},\nonumber
\end{align}
where $\omega_{cj}=q_{j}B_{0}/m_{j}c$ is the cyclotron frequency. The equation
for the perturbed longitudinal velocity $v_{j1z}$ is given by%
\begin{align}
\left(  D_{tj}^{2}-\frac{T_{j0}}{m_{j}}\frac{\partial^{2}}{\partial z^{2}%
}\right)  v_{j1z}  &  =-\frac{1}{m_{j}}D_{tj}\frac{\partial T_{j1}}{\partial
z}+D_{tj}F_{j1z},\tag{A4}\\
F_{e1z}  &  =\frac{q_{e}}{m_{e}}E_{1z}-\nu_{ei0}v_{e1z}+\nu_{ei0}\frac{D_{te}%
}{D_{ti}}v_{i1z},\nonumber\\
F_{i1z}  &  =\frac{q_{i}}{m_{i}}E_{1z}-\nu_{ie0}v_{i1z}+\nu_{ie0}\frac{D_{ti}%
}{D_{te}}v_{e1z},\nonumber
\end{align}
where we have used the linearized continuity equation (2).
\bigskip
\subsection{Perturbed temperatures of ions and electrons}
From the linearized equations (3) and (4), we obtain equations for the
perturbed temperatures of ions and electrons, $T_{i,e1}$. We will assume that
the background ion and electron temperatures are equal to each other,
$T_{i0}=T_{e0}=T_{0}$. In this case, the terms connected with the perturbation
of thermal energy exchange frequency in equations (3) and (4) will be absent.
However for convenience of calculations to follow the symmetric contribution
of ions and electrons, we formally retain different notations for the ion and
electron temperatures. Then, we will have%
\begin{align}
D_{i}T_{i1}-\Omega_{ie}T_{e1}  &  =-\left(  \gamma-1\right)  T_{i0}%
\frac{\partial v_{i1z}}{\partial z},\tag{A5}\\
D_{e}T_{e1}-\Omega_{ei}T_{i1}  &  =-\left(  \gamma-1\right)  T_{e0}%
\frac{\partial v_{e1z}}{\partial z}.\nonumber
\end{align}
Here, the following notations are introduced:%
\begin{align}
D_{i}  &  =D_{ti}+\Omega_{ie},D_{e}=D_{te}+\Omega_{\chi}+\Omega_{ei}%
,\tag{A6}\\
\Omega_{ie}  &  =\nu_{ie}^{\varepsilon}\left(  n_{e0},T_{e0}\right)
,\Omega_{ei}=\nu_{ei}^{\varepsilon}\left(  n_{i0},T_{e0}\right)  ,\nonumber\\
\Omega_{\chi}  &  =-\left(  \gamma-1\right)  \frac{\chi_{e0}}{n_{e0}}%
\frac{\partial^{2}}{\partial z^{2}},\nonumber
\end{align}
where we have used equation (5) for obtaining $\Omega_{\chi}$. Solutions of
equation (A5) for $T_{i,e1}$ are given by%
\begin{align}
DT_{i1}  &  =-D_{e}\left(  \gamma-1\right)  T_{i0}\frac{\partial v_{i1z}%
}{\partial z}-\Omega_{ie}\left(  \gamma-1\right)  T_{e0}\frac{\partial
v_{e1z}}{\partial z},\tag{A7}\\
DT_{e1}  &  =-D_{i}\left(  \gamma-1\right)  T_{e0}\frac{\partial v_{e1z}%
}{\partial z}-\Omega_{ei}\left(  \gamma-1\right)  T_{i0}\frac{\partial
v_{i1z}}{\partial z},\nonumber
\end{align}
where%
\begin{equation}
D=D_{i}D_{e}-\Omega_{ie}\Omega_{ei}. \tag{A8}%
\end{equation}
\bigskip
\subsection{Equations for longitudinal velocities $v_{i1z}$\ and $v_{e1z}$}
Let us substitute equation (A7) into equation (A4) written for the ions and
electrons. Then, we obtain%
\begin{align}
L_{1i}v_{i1z}+L_{2i}v_{e1z}  &  =DD_{ti}\frac{q_{i}}{m_{i}}E_{1z},\tag{A9}\\
L_{1e}v_{e1z}+L_{2e}v_{i1z}  &  =DD_{te}\frac{q_{e}}{m_{e}}E_{1z}.\nonumber
\end{align}
Here, we have introduced notations%
\begin{align}
L_{1i}  &  =DD_{ti}\left(  D_{ti}+\nu_{ie0}\right)  -\frac{T_{i0}}{m_{i}%
}D\frac{\partial^{2}}{\partial z^{2}}-\left(  \gamma-1\right)  \frac{T_{i0}%
}{m_{i}}D_{ti}D_{e}\frac{\partial^{2}}{\partial z^{2}},\tag{A10}\\
L_{1e}  &  =DD_{te}\left(  D_{te}+\nu_{ei0}\right)  -\frac{T_{e0}}{m_{e}%
}D\frac{\partial^{2}}{\partial z^{2}}-\left(  \gamma-1\right)  \frac{T_{e0}%
}{m_{e}}D_{te}D_{i}\frac{\partial^{2}}{\partial z^{2}},\nonumber\\
L_{2i}  &  =-D_{ti}\left[  \left(  \gamma-1\right)  \frac{T_{e0}}{m_{i}}%
\Omega_{ie}\frac{\partial^{2}}{\partial z^{2}}+\nu_{ie0}D\frac{D_{ti}}{D_{te}%
}\right]  ,\nonumber\\
L_{2e}  &  =-D_{te}\left[  \left(  \gamma-1\right)  \frac{T_{i0}}{m_{e}}%
\Omega_{ei}\frac{\partial^{2}}{\partial z^{2}}+\nu_{ei0}D\frac{D_{te}}{D_{ti}%
}\right]  .\nonumber
\end{align}
Solutions of equation (A9) are the following:%
\begin{align}
v_{i1z}  &  =\frac{D}{L}\left(  L_{1e}D_{ti}\frac{q_{i}}{m_{i}}-L_{2i}%
D_{te}\frac{q_{e}}{m_{e}}\right)  E_{1z},\tag{A11}\\
v_{e1z}  &  =\frac{D}{L}\left(  L_{1i}D_{te}\frac{q_{e}}{m_{e}}-L_{2e}%
D_{ti}\frac{q_{i}}{m_{i}}\right)  E_{1z},\nonumber
\end{align}
where
\begin{equation}
L=L_{1i}L_{1e}-L_{2i}L_{2e}. \tag{A12}%
\end{equation}
\bigskip
\subsection{Expressions for perturbed transverse velocities via $\mathbf{E}%
_{1}$\ }
Using equation (9), we can find the components of $\mathbf{F}_{j1}$ given by
equation (A2). In the case under consideration, we obtain%
\begin{equation}
F_{j1x,y}=\frac{q_{j}}{m_{j}}D_{tj}\left(  \frac{\partial}{\partial t}\right)
^{-1}E_{1x,y}. \tag{A13}%
\end{equation}
Substitution of equation (A13) into equation (A3) gives
\begin{align}
\left(  D_{tj}^{2}+\omega_{cj}^{2}\right)  v_{j1x}  &  =\frac{q_{j}}{m_{j}%
}\omega_{cj}D_{tj}\left(  \frac{\partial}{\partial t}\right)  ^{-1}%
E_{1y}+\frac{q_{j}}{m_{j}}D_{tj}^{2}\left(  \frac{\partial}{\partial
t}\right)  ^{-1}E_{1x},\tag{A14}\\
\left(  D_{tj}^{2}+\omega_{cj}^{2}\right)  v_{j1y}  &  =-\frac{q_{j}}{m_{j}%
}\omega_{cj}D_{tj}\left(  \frac{\partial}{\partial t}\right)  ^{-1}%
E_{1x}+\frac{q_{j}}{m_{j}}D_{tj}^{2}\left(  \frac{\partial}{\partial
t}\right)  ^{-1}E_{1y}.\nonumber
\end{align}
\bigskip
\subsection{Perturbed plasma current}
The components of the transverse perturbed plasma current $j_{pl1x,y}=\sum
_{j}q_{j}n_{j0}v_{j1x,y}$ are found by using equation (A14). The expressions
for $4\pi\left(  \partial/\partial t\right)  ^{-1}j_{pl1x,y}$ can be given in
the form%
\begin{align}
4\pi\left(  \frac{\partial}{\partial t}\right)  ^{-1}j_{pl1x}  &
=\varepsilon_{plxx}E_{1x}+\varepsilon_{plxy}E_{1y},\tag{A15}\\
4\pi\left(  \frac{\partial}{\partial t}\right)  ^{-1}j_{pl1y}  &
=-\varepsilon_{plxy}E_{1x}+\varepsilon_{plxx}E_{1y},\nonumber
\end{align}
where $\omega_{pj}=\left(  4\pi n_{j0}q_{j}^{2}/m_{j}\right)  ^{1/2}$ is the
plasma frequency. The following notations are introduced in equation (A15):%
\begin{align}
\varepsilon_{plxx}  &  =\sum_{j}\frac{\omega_{pj}^{2}D_{tj}^{2}}{\left(
D_{tj}^{2}+\omega_{cj}^{2}\right)  }\left(  \frac{\partial}{\partial
t}\right)  ^{-2},\tag{A16}\\
\varepsilon_{plxy}  &  =\sum_{j}\frac{\omega_{pj}^{2}\omega_{cj}D_{tj}%
}{\left(  D_{tj}^{2}+\omega_{cj}^{2}\right)  }\left(  \frac{\partial}{\partial
t}\right)  ^{-2}.\nonumber
\end{align}
The longitudinal perturbed plasma current $j_{pl1z}=\sum_{j}q_{j}n_{j0}%
v_{j1z}+\sum_{j}q_{j}n_{j1}v_{j0}$ is found by using equations (2) and (A11)%
\begin{align}
4\pi\left(  \frac{\partial}{\partial t}\right)  ^{-1}j_{pl1z}  &
=\frac{\omega_{pi}^{2}}{D_{ti}}\frac{D}{L}\left(  L_{1e}D_{ti}-L_{2i}%
D_{te}\frac{q_{e}m_{i}}{q_{i}m_{e}}\right)  E_{1z}\tag{A17}\\
&  +\frac{\omega_{pe}^{2}}{D_{te}}\frac{D}{L}\left(  L_{1i}D_{te}-L_{2e}%
D_{ti}\frac{q_{i}m_{e}}{q_{e}m_{i}}\right)  E_{1z}.\nonumber
\end{align}
\bigskip
\section{Appendix}
\subsection{Perturbed velocity of cosmic rays}
The linearized version of equation (6) for $\mathbf{p}_{cr1}=$ $\mathbf{p}%
_{cr}-\mathbf{p}_{cr0}$ and $\mathbf{v}_{cr0}\parallel\mathbf{z}$ has the form%
\begin{equation}
R_{cr0}D_{tcr}\mathbf{p}_{cr1}+\mathbf{p}_{cr0}D_{tcr}R_{cr1}=-\frac
{\mathbf{\nabla}p_{cr1}}{n_{cr0}}+m_{cr}\mathbf{F}_{cr1}\mathbf{+}\frac
{q_{cr}}{c}\mathbf{v}_{cr1}\times\mathbf{B}_{0}, \tag{B1}%
\end{equation}
where $D_{tcr}=\partial/\partial t+v_{cr0}\partial/\partial z$ and
\begin{equation}
m_{cr}\mathbf{F}_{cr1}=q_{cr}\mathbf{E}_{1}\mathbf{+}\frac{q_{cr}}%
{c}\mathbf{v}_{cr0}\times\mathbf{B}_{1}. \tag{B2}%
\end{equation}
For the perturbed transverse velocities of cosmic rays $v_{cr1x,y}$, we find
from equation (B1) the following solutions:%
\begin{align}
\left(  D_{cr}^{2}+\omega_{ccr}^{2}\right)  v_{cr1x}  &  =\frac{q_{cr}}%
{m_{cr}}\omega_{ccr}D_{tcr}\left(  \frac{\partial}{\partial t}\right)
^{-1}E_{1y}+\frac{q_{cr}}{m_{cr}}D_{cr}D_{tcr}\left(  \frac{\partial}{\partial
t}\right)  ^{-1}E_{1x},\tag{B3}\\
\left(  D_{cr}^{2}+\omega_{ccr}^{2}\right)  v_{cr1y}  &  =-\frac{q_{cr}%
}{m_{cr}}\omega_{ccr}D_{tcr}\left(  \frac{\partial}{\partial t}\right)
^{-1}E_{1x}+\frac{q_{cr}}{m_{cr}}D_{cr}D_{tcr}\left(  \frac{\partial}{\partial
t}\right)  ^{-1}E_{1y},\nonumber
\end{align}
where $D_{cr}=\gamma_{cr0}R_{cr0}D_{tcr}$. When obtaining equation (B3), we
have expressed $F_{cr1x,y}$ through $E_{1x,y}$ by using equation (A13) for
$j=cr$ (see equation (B2)).
The $z$-component of equation (B1) is given by
\begin{equation}
D_{cr}v_{cr1z}+v_{cr0}R_{cr0}D_{tcr}\gamma_{cr1}+\gamma_{cr0}v_{cr0}%
D_{tcr}R_{cr1}=-\frac{1}{m_{cr}n_{cr0}}\frac{\partial p_{cr1}}{\partial
z}+F_{cr1z}, \tag{B4}%
\end{equation}
where $\gamma_{cr1}=\gamma_{cr0}^{3}v_{cr0}v_{cr1z}/c^{2}$.
\bigskip
\subsection{Perturbed temperature and pressure of cosmic rays}
We now find $R_{cr1}$ and $p_{cr1}$. From equation (8), we see that
\begin{equation}
R_{cr1}=\frac{\Gamma_{cr}}{\Gamma_{cr}-1}\frac{T_{cr1}}{m_{cr}c^{2}}. \tag{B5}%
\end{equation}
The perturbation of the temperature $T_{cr1}$ found from the equation
$T_{cr}=\gamma_{cr}p_{cr}/n_{cr}$ is equal to%
\begin{equation}
T_{cr1}=T_{cr0}\left(  \frac{p_{cr1}}{p_{cr0}}-\frac{n_{cr1}}{n_{cr0}}%
+\frac{\gamma_{cr1}}{\gamma_{cr0}}\right)  . \tag{B6}%
\end{equation}
From equation (7), we can find the pressure perturbations $p_{cr1}$%
\begin{equation}
p_{cr1}=p_{cr0}\Gamma_{cr}\left(  \frac{n_{cr1}}{n_{cr0}}-\frac{\gamma_{cr1}%
}{\gamma_{cr0}}\right)  , \tag{B7}%
\end{equation}
where%
\begin{equation}
n_{cr1}=-n_{cr0}\frac{\partial v_{cr1z}}{D_{tcr}\partial z}. \tag{B8}%
\end{equation}
\bigskip
\subsection{Equation for\textbf{ }$v_{cr1z}$}
Substituting equations (B5)-(B8) into equation (B4) and using equation (A13)
for cosmic rays, we obtain%
\begin{equation}
\gamma_{cr0}^{3}\left(  R_{cr0}-\frac{\Gamma_{cr}T_{cr0}}{m_{cr}c^{2}}%
\frac{v_{cr0}^{2}}{c^{2}}\right)  D_{tcr}^{2}v_{cr1z}-2\gamma_{cr0}%
v_{cr0}\frac{\Gamma_{cr}T_{cr0}}{m_{cr}c^{2}}D_{tcr}\frac{\partial v_{cr1z}%
}{\partial z}-\frac{\Gamma_{cr}T_{cr0}}{\gamma_{cr0}m_{cr}}\frac{\partial
^{2}v_{cr1z}}{\partial z^{2}}=D_{tcr}\frac{q_{cr}}{m_{cr}}E_{1z}. \tag{B9}%
\end{equation}
\bigskip
\subsection{Perturbed cosmic-ray current}
The perturbed transverse cosmic ray currents $j_{cr1x,y}=q_{cr}n_{cr0}%
v_{cr1x,y}$ are found by using equation (B3). For the values $4\pi\left(
\partial/\partial t\right)  ^{-1}j_{cr1x,y}$, we will have%
\begin{align}
4\pi\left(  \frac{\partial}{\partial t}\right)  ^{-1}j_{cr1x}  &
=\varepsilon_{crxx}E_{1x}+\varepsilon_{crxy}E_{1y},\tag{B10}\\
4\pi\left(  \frac{\partial}{\partial t}\right)  ^{-1}j_{cr1y}  &
=-\varepsilon_{crxy}E_{1x}+\varepsilon_{crxx}E_{1y}.\nonumber
\end{align}
Here%
\begin{align}
\varepsilon_{crxx}  &  =\frac{\omega_{pcr}^{2}D_{cr}D_{tcr}}{\left(
D_{cr}^{2}+\omega_{ccr}^{2}\right)  }\left(  \frac{\partial}{\partial
t}\right)  ^{-2},\tag{B11}\\
\varepsilon_{crxy}  &  =\frac{\omega_{pcr}^{2}\omega_{ccr}D_{tcr}}{\left(
D_{cr}^{2}+\omega_{ccr}^{2}\right)  }\left(  \frac{\partial}{\partial
t}\right)  ^{-2},\nonumber
\end{align}
where $\omega_{pcr}=\left(  4\pi n_{cr0}q_{cr}^{2}/m_{cr}\right)  ^{1/2}$ is
the cosmic ray plasma frequency.
The longitudinal perturbed cosmic ray current is equal to $j_{cr1z}%
=q_{cr}n_{cr0}v_{cr1z}+q_{cr}n_{cr1}v_{cr0}$. Making use of the linearized
continuity equation for cosmic rays and equation (B9), we find%
\begin{equation}
4\pi\left(  \frac{\partial}{\partial t}\right)  ^{-1}j_{cr1z}=\frac
{\omega_{pcr}^{2}}{L_{cr}}E_{1z}, \tag{B12}%
\end{equation}
where%
\begin{equation}
L_{cr}=\gamma_{cr0}^{3}\left(  R_{cr0}-\frac{\Gamma_{cr}T_{cr0}}{m_{cr}c^{2}%
}\frac{v_{cr0}^{2}}{c^{2}}\right)  D_{tcr}^{2}-2\gamma_{cr0}v_{cr0}%
\frac{\Gamma_{cr}T_{cr0}}{m_{cr}c^{2}}D_{tcr}\frac{\partial}{\partial z}%
-\frac{\Gamma_{cr}T_{cr0}}{\gamma_{cr0}m_{cr}}\frac{\partial^{2}}{\partial
z^{2}}. \tag{B13}%
\end{equation}
\bigskip
\end{appendix}

\end{document}